\documentclass[12pt]{article}

\newcommand{\q}{\underline{q}}
\newcommand{\sq}{\scriptstyle{\underline{q}}}

\usepackage{graphicx}
\begin{document}

\title{
{\sl k}-Component $q$-deformed charge coherent states
and their nonclassical properties
}
\author{{X.-M. Liu$~^{{\tt a,b,*}}$, C. Quesne$~^{{\tt b}}$, F. Song$~^{{\tt a}}$} \\
{\small ${}^{{\tt a}}$ Department of Physics, Beijing Normal University,
Beijing 100875, China ${}^1$ }\\
{\small ${}^{{\tt b}}$ Physique Nucl\'{e}aire Th\'{e}orique et Physique
Math\'{e}matique,}\\
{\small  Universit\'{e} Libre de Bruxelles, Campus de la Plaine CP229,}\\
{\small Boulevard du Triomphe, B-1050 Brussels, Belgium }}
\date{}
\maketitle


\begin{center}
\begin{minipage}{120mm}
\vskip 0.12in
\baselineskip 0.374in
\begin{center}{\bf Abstract}\end{center}
\mbox{}\hspace{6mm}{\sl k}-Component $q$-deformed charge coherent states are
constructed, their (over)completeness proved and their generation explored.
The $q$-deformed charge coherent states and the even (odd) $q$-deformed charge
coherent states are the two special cases of them as $k$ becomes 1 and 2, respectively.
A $D$-algebra realization of the SU$_q$(1,1) generators is given in terms of them.
Their nonclassical properties are studied and it is shown that for $k\geq3$, they exhibit two-mode $q$-antibunching, but neither
SU$_q$(1,1) squeezing, nor one- or two-mode $q$-squeezing.
\vskip 0.20in
Keywords: {\sl k}-Component charge coherent states; $q$-Deformation; Completeness relations;
Squeezing

PACS number(s): 42.50.Dv; 03.65.-W; 03.65.Ca; 03.65.Fd
\end{minipage}
\end{center}

\vskip 1mm 

{\small ${}^*$ Corresponding author}

{\small ~~E-mail addresses: liuxm@bnu.edu.cn (X.-M. Liu),
cquesne@ulb.ac.be (C. Quesne). }

{\small ${}^1$ Mailing address} 


\newpage

\baselineskip 0.374in

\section*{1. Introduction}

\mbox{}\hspace{6mm}The coherent states introduced by Schr\"{o}dinger [1] and
Glauber [2] are the eigenstates of the boson annihilation operator, and have
widespread applications in the fields of physics [3$-$7]. However, in all
the cases the quanta involved are uncharged. In 1976, Bhaumik et al.
[4,8,9] constructed the boson coherent states which, carrying definite charge,
are the eigenstates of both the pair boson annihilation operator and the
charge operator. This kind of states are the so-called charge coherent
states. Based on this work, the charge coherent states for SU(2) [10],
SU(3) [11], and arbitrary compact Lie groups [12] were also put forward.

The concept of charge coherent states has proved to be very useful in many
areas, such as elementary particle physics [9,13$-$17], quantum field theory
[12,18,19], nuclear physics [20], thermodynamics [21$-$23], quantum
mechanics [24], and quantum optics [25$-$27]. Moreover, some schemes for
generating charge coherent states in quantum optics were proposed
[25,26,28,29].

As is well known, the even and odd coherent states [30], which are the two
orthonormalized eigenstates of the square of the boson annihilation
operator, play an important role in quantum optics [31$-$33].
An extension of the even and odd coherent states is to define the
{\sl k}-component coherent states [34,35], which are the $k$ $(k\geq 1)$
orthonormalized eigenstates of the $k$th power of the boson annihilation
operator. The coherent states and the even (odd) ones are the two
special cases of the {\sl k}-component coherent states as $k$ becomes 1 and
2, respectively. Inspired by the above idea, in Ref. [36] one of the authors
(X.-M.L.) has generalized the charge coherent states to the even and odd
charge coherent states, defined as the two orthonormalized eigenstates of
both the square of the pair boson annihilation operator and the charge
operator; in Ref. [37] he has further extended the even and odd charge
coherent states to the {\sl k}-component charge coherent states, defined as
the $k$ orthonormalized eigenstates of both the $k$th power of the pair boson
annihilation operator and the charge operator. The charge coherent states and
the even (odd) ones are the two special cases of the {\sl k}-component charge
coherent states as $k$ becomes 1 and 2, respectively.

On the other hand, quantum groups [38,39], introduced as a mathematical
description of deformed Lie algebras, have given the possibility of
generalizing the notion of coherent states to the case of $q$-deformations
[40$-$44]. A $q$-deformed harmonic oscillator [40,45] was defined in terms
of $q$-boson annihilation and creation operators, the latter satisfying the
quantum Heisenberg-Weyl algebra [40,45,46], which plays an important role in
quantum groups. The $q$-deformed coherent states introduced by Biedenharn
[40] are the eigenstates of the $q$-boson annihilation operator. Such states
have been well studied [41,42,47,48], and widely applied to quantum optics
and mathematical physics [44,49$-$53]. Furthermore, the $q$-deformed charge
coherent states [54,55] were constructed as the eigenstates of both the
pair $q$-boson annihilation operator and the charge operator.

A natural extension of the $q$-deformed coherent states is provided by the
even and odd $q$-deformed coherent states [56], which are the two
orthonormalized eigenstates of the square of the $q$-boson annihilation
operator. In a previous Letter [57], motivated by the above idea, the authors
(X.-M.L. and C.Q.)
have generalized the $q$-deformed charge coherent states to the even and odd
$q$-deformed charge coherent states, defined as the two orthonormalized
eigenstates of both the square of the pair $q$-boson annihilation operator
and the charge operator.
A further extension of the even and odd $q$-deformed coherent states is
given by the {\sl k}-component $q$-deformed coherent states [58,59], which
are the $k$ orthonormalized eigenstates of the $k$th power of the $q$-boson
annihilation operator. The $q$-deformed coherent states and the even (odd)
ones are the two special cases of the {\sl k}-component $q$-deformed
coherent states as $k$ becomes 1 and 2, respectively. In a parallel way, it
is very desirable to generalize the even and odd $q$-deformed charge
coherent states to the {\sl k}-component $q$-deformed charge coherent states,
defined as the $k$ orthonormalized eigenstates of both the $k$th power of the
pair $q$-boson annihilation operator and the charge operator. The
$q$-deformed charge coherent states and the even (odd) ones are the two
special cases of the {\sl k}-component $q$-deformed charge coherent states as
$k$ becomes 1 and 2, respectively.

This paper is arranged as follows. In Section 2, the {\sl k}-component
$q$-deformed charge coherent states are constructed. Their completeness is
proved in Section 3. Section 4 is devoted to generating them. In Section 5,
they are used to provide a $D$-algebra realization of the SU$_q(1,1)$
generators. Their nonclassical properties, such as SU$_q(1,1)$ squeezing,
single- or two-mode $q$-squeezing, and two-mode $q$-antibunching, are
studied in Section 6. Section 7 contains a summary of the results.


\section*{2. \boldmath {\sl k}-Component $q$-deformed charge coherent states}

\mbox{}\hspace{6mm}Two mutually commuting $q$-deformed harmonic
oscillators are defined in terms of two pairs of independent $q$-boson
annihilation and creation operators $a_i$, $a_i^{+}$ $(i=1,2)$, together
with corresponding number operators $N_i$, satisfying the quantum
Heisenberg-Weyl algebra
\begin{equation}  \label{e1}
{a_i}a_i^{+}-qa_i^{+}{a_i}=q^{-N_i},
\end{equation}
\begin{equation}  \label{e2}
[{N_i},~a_i^{+}]=a_i^{+},\qquad[{N_i},~{a_i}]=-{a_i},
\end{equation}
where $q$ is a positive real deformation parameter.
The operators $a_i$, $a_i^{+}$, and $N_i$ act in the Fock space with basis $
|n\rangle_i$ $(n=0,1,2,\ldots )$, such that
\begin{equation}  \label{e3}
{a_i}|0{\rangle}_i=0,\qquad |n{\rangle}_i={\frac{{(a_i^{+})}^n}{%
\sqrt{[n]!}}} |0{\rangle}_i,
\end{equation}
where
\begin{equation}  \label{e4}
[n]!{\equiv}[n]_q!=[n]_q[n-1]_q\ldots [1]_q,\qquad [0]!=1,
\end{equation}
\begin{equation}  \label{e5}
[n]_q=\frac{q^n-q^{-n}}{q-q^{-1}}{\equiv}[n].
\end{equation}
Their action on the basis states is given by
\begin{equation}  \label{e6}
{a_i}|n{\rangle}_i=\sqrt{[n]}|n-1{\rangle} _i,\qquad {a_i^{+}}|n{%
\rangle}_i={\sqrt{[n+1]}}|n+1{\rangle} _i,\qquad {N_i}|n{\rangle}%
_i=n|n{\rangle}_i.
\end{equation}
Note that $[n]$ is invariant under $q~{\leftrightarrow}~1/q$.
In the following, $[n]$ will refer to the $q$-deformed  $n$ defined by
(\ref{e5}) corresponding to the base $q$. If the base is different, then it will be
indicated explicitly.

The $q$-boson operators $a_i$ and $a_i^{+}$ can be constructed from the
conventional boson annihilation and creation operators $b_i$, $b_i^{+}$ in
the following way [60]:
\begin{equation}  \label{e7}
{a_i}=\sqrt{\frac{[N_i+1]}{N_i+1}}b_i,\qquad {a_i^+}={b_i^+}\sqrt{%
\frac{[N_i+1]}{N_i+1}},
\end{equation}
where $N_i=b_i^{+}b_i$. It is worth noticing that $[N_i]=a_i^{+}a_i$.

The operators $a_1(a_1^{+})$ and $a_2(a_2^{+})$ are assigned the ``charge''
quanta 1 and $-1$, respectively. Thus the charge operator is given by
\begin{equation}
Q=N_1-N_2.  \label{e8}
\end{equation}
In view of the fact that
\begin{equation}
\lbrack Q,{(a_1a_2)}^k]=0,  \label{e9}
\end{equation}
where $k$ is a positive integer $(k=1,2,3,...)$, we may seek the
{\sl k}-component $q$-deformed charge coherent states, which are
the $k$ orthonormalized eigenstates of both the $k$th power
${(a_1a_2)}^k$ of the pair $q$-boson annihilation operator ${a_1a_2}$
and the charge operator $Q$.

Let $|m,n{\rangle }=|m{\rangle }_1|n{\rangle }_2$ denote the basis states of
two-mode Fock space, where $|m{\rangle }_1$ and $|n{\rangle }_2$ are the
eigenstates of $N_1$ and $N_2$ corresponding to the eigenvalues $m$ and $n$,
respectively. They satisfy the completeness relation
\begin{equation}
\sum\limits_{m=0}^\infty \sum\limits_{n=0}^\infty |m,n{\rangle }{\langle }%
m,n|=I.
\label{e10}
\end{equation}
We now consider the following states:
\begin{eqnarray}
|{\xi },\q,k{\rangle }_j&=&N_{k\sq}^j{\ \sum_{p=\max(0,-\q/k%
)}^\infty }\frac{{\xi }^{kp+j+\min(0,\sq)}}{{\{[kp+j]![kp+j+\q]!\}}%
^{1/2}}|kp+j+\q,kp+j{\rangle } \nonumber \\
&=&\left\{
\begin{array}{ll}
N_{k\sq}^j{\sum\limits_{n=0}^\infty }\frac{{\xi }^{kn+j}}{{\{[kn+j]![kn+j+%
\sq]!\}}^{1/2}}|kn+j+\q,kn+j{\rangle }, & \q\geq 0,
\\
N_{k\sq}^j{\sum\limits_{n=0}^\infty }\frac{{\xi }^{kn+j}}{{\{[kn+j]![kn+j-%
\sq]!\}}^{1/2}}|kn+j,kn+j-\q{\rangle }, & \q\leq 0,
\end{array}
\right.  \label{e11}
\end{eqnarray}
where $j=0,1,\ldots ,k-1$, $\xi $ is a complex number, $\q$ is a fixed
integer and $N_{k\sq}^j$ are normalization factors given by
\begin{equation}
N_{k\sq}^j{\equiv }N_{k\sq}^j({|\xi |}^2)={ \left\{
\sum\limits_{n=0}^\infty \frac{{({|\xi |}^2)}^{kn+j}}{[kn+j]![kn+j+|\q|]!}%
\right\} }^{-{1/2}}.  \label{e12}
\end{equation}
As can be verified by explicit calculations, these states satisfy the
relations
\begin{equation}
{(a_1a_2)}^k|{\xi },\q,k{\rangle }_{j}={\xi }^k|{\xi },\q,k{\rangle }%
_{j}, \qquad Q|{\xi },\q,k{\rangle }_{j}=\q|{\xi },\q,k{%
\rangle }_{j}, \qquad {}_j{\langle }\xi ,\q,k|\xi ,\q,k{\rangle }%
_{j^{\prime }}={\delta }_{jj{\prime }}.  \label{e13}
\end{equation}
It indicates that $|{\xi },\q,k{\rangle }_j$ ($j=0,1,\ldots ,k-1$) in (11)
are exactly the $k$ orthonormalized eigenstates of both the operator
${(a_1a_2)}^k$ and $Q$ corresponding to the eigenvalues ${\xi }^k$ and $\q$,
respectively. Obviously, $\q$ is the charge number which the states
$|{\xi },\q,k{\rangle }_j $ carry.

Therefore, the $k$ states of (11) are just what we want, that is to say,
they are the {\sl k}-component $q$-deformed charge coherent states.
In the limit $q{\rightarrow}1$, they reduce to the usual {\sl k}-component
charge coherent states constructed by the author (X.-M.L.) [37].

According to (11), for $k=1$, we obtain
\begin{eqnarray}
|{\xi },\q,1{\rangle }_0&=&N_{\sq}{\ \sum_{p=\max(0,-\sq)}^\infty }\frac{{%
\xi }^{p+\min(0,\sq)}}{{\{[p]![p+\q]!\}}^{1/2}}|p+\q,p{\rangle }
\nonumber \\
&=&\left\{
\begin{array}{ll}
N_{\sq}{\sum\limits_{n=0}^\infty }\frac{{\xi }^n}{{\{[n]![n+\sq]!\}}
^{1/2}}|n+\q,n{\rangle }, &\q\geq 0, \\[0.2cm]
N_{\sq}{\sum\limits_{n=0}^\infty }\frac{{\xi }^n}{{\{[n]![n-\sq]!\}}
^{1/2}}|n,n-\q{\rangle }, & \q\leq 0,
\end{array}
\right.
\nonumber \\
&\equiv&|{\xi },\q{\rangle },
\label{e14}
\end{eqnarray}
where
\begin{equation}
N_{\sq}{\equiv }N_{1\sq}^0({|\xi |}^2)={ \left\{ \sum\limits_{n=0}^\infty
\frac{{({|\xi |}^2)}^n}{[n]![n+|\q|]!}\right\} }^{-{1/2}}.
\label{e15}
\end{equation}
It is evident that $|{\xi },\q,1{\rangle }_0({\equiv }|{\xi },\q{\rangle })$
are exactly the so called $q$-deformed charge coherent states
given in Ref. [54].

According to (11), for $k=2$, we obtain
\begin{eqnarray}
|{\xi },\q,2{\rangle }_j&=&N_{2\sq}^j{\ \sum_{p=\max(0,-\q/2%
)}^\infty }\frac{{\xi }^{2p+j+\min(0,\sq)}}{{\{[2p+j]![2p+j+\q]!\}}%
^{1/2}}|2p+j+\q,2p+j{\rangle } \nonumber \\
&=&\left\{
\begin{array}{ll}
N_{2\sq}^j{\sum\limits_{n=0}^\infty }\frac{{\xi }^{2n+j}}{{\{[2n+j]![2n+j+%
\sq]!\}}^{1/2}}|2n+j+\q,2n+j{\rangle }, & \q\geq 0,
\\
N_{2\sq}^j{\sum\limits_{n=0}^\infty }\frac{{\xi }^{2n+j}}{{\{[2n+j]![2n+j-%
\sq]!\}}^{1/2}}|2n+j,2n+j-\q{\rangle }, & \q\leq 0,
\end{array}
\right.  \label{e16}
\end{eqnarray}
where $j=0,1$. It is evident that $|{\xi },\q,2{\rangle }_0$~$(|{\xi },\q,2{%
\rangle }_1)$ are exactly the so called even (odd) $q$-deformed charge
coherent states obtained by the authors (X.-M.L. and C.Q.) [57].

From (11), it follows that
\begin{equation}
{}_{j}{\langle }\xi ,\q,k|{\xi }^{\prime },\q^{\prime },k{\rangle }%
_{j^{\prime }}=N_{k\sq}^{j}({|\xi |}^2)N_{k\sq}^{j}({|{\xi }^{\prime }|}^2){%
\left[ N_{k\sq}^{j}({\xi }^{*}{\xi }^{\prime })\right] }^{-2}{\delta }_{%
\q\q'}{\delta }_{jj^{\prime }}. \label{e17}
\end{equation}
This further shows that, for the same value of $k$, the states
$|{\xi },\q,k{\rangle }_j$ are orthogonal to one another with respect to
both the subscript $j$ and the charge number $\q$. However, they are
nonorthogonal with respect to the parameter $\xi$.

For the mean values of the operators $N_1$ and $N_2$, there exists the
relation
\begin{equation}
{}_{j}{\langle }\xi ,\q,k|N_1|{\xi },\q,k{\rangle }_{j}=\q + {}_{j}%
{\langle }\xi ,\q,k|N_2|\xi,\q,k{\rangle }_{j}.  \label{e18}
\end{equation}

In terms of the {\sl k}-component $q$-deformed charge coherent states, the
$q$-deformed charge coherent states can be expanded as
\begin{equation}
|\xi ,\q{\rangle }=N_{\sq}\left[ \sum\limits_{j=0}^{k-1}
{(N_{k\sq}^j)}^{-1}|\xi ,\q,k{\rangle }_j\right],  \label{e19}
\end{equation}
where the normalization factors are such that
\begin{equation}
N_{\sq}^{-2}   =\sum\limits_{j=0}^{k-1}{(N_{k\sq}^j)}^{-2}.  \label{e20}
\end{equation}

\section*{\boldmath 3. Completeness of {\sl k}-component $q$-deformed charge
coherent states }

\mbox{}\hspace{6mm}Let us begin with some $q$-deformed formulas which are
useful in the proof of completeness of the {\sl k}-component $q$-deformed
charge coherent states. The $q$-deformed Bessel function of (integer) order
$\nu $ may be defined by~[61]
\begin{equation}
J_\nu (q,x)=\sum\limits_{k=0}^\infty \frac{{(-1)}^k}{[k]![\nu
+k]!}{\left( \frac x{\sqrt{q}[2]_{\sqrt{q}}}\right) }^{\nu +2k},  \label{e21}
\end{equation}
where $[n]_{\sqrt{q}}$ is defined as in Eq. (5) except for replacing
$q$ by $\sqrt{q}$. An integral representation of the
$q$-deformed modified Bessel function of order $\nu$ is given by~[62]
\begin{equation}
K_\nu (q,x)=\frac 1{[2]_{\sqrt{q}}}{\left( \frac x{[2]_{\sqrt{q}}}\right) }%
^\nu \int\limits_0^\infty d_qt\frac 1{t^{\nu +1}}e_q(-t)e_q\left( -\frac{x^2%
}{{([2]_{\sqrt{q}})}^2t}\right),  \label{e22}
\end{equation}
where $d_qt$ is a standard $q$-integration [47,63,64], and $e_q(x)$ is a $q$%
-exponential function~[47]
\begin{equation}
e_q(x)=\left\{
\begin{array}{ll}
{\sum\limits_{n=0}^\infty }\frac{x^n}{[n]!}, &{\rm for\ }x>-{\zeta }, \\[0.2cm]
0, & {\rm otherwise},
\end{array}
\right.  \label{e23}
\end{equation}
with $-\zeta $ being the largest zero of $e_q(x)$. Then, it follows that [62]
\begin{equation}
\int\limits_0^\infty d_{\sqrt{q}}u\,u^{2p+{\nu }+1}K_\nu (q,[2]_{\sqrt{q}}u)=%
\frac{[\nu +p]![p]!}{{([2]_{\sqrt{q}})}^2}.  \label{e24}
\end{equation}

We now prove that the {\sl k}-component $q$-deformed charge coherent states
form an (over)complete set, that is to say
\begin{equation}
\sum\limits_{\sq={-\infty }}^\infty \int \frac{d_{q}^2{\xi }}\pi {\phi }_{%
\sq}(\xi )N_{\sq}^2\left[\sum\limits_{j=0}^{k-1}{(N_{k\sq}^j)}^{-2}
|\xi ,\q,k{\rangle}_j\,{}_j{\langle }\xi ,\q,k|\right]
\equiv \sum\limits_{\sq={-\infty }}^\infty I_{\sq}=I,
\label{e25}
\end{equation}
where
\begin{equation}
d_{q}^2{\xi }=|\xi|d_{\sqrt{q}}|\xi|d{\theta}, \qquad {\xi}%
=|\xi|e^{i\theta},  \label{e26}
\end{equation}
and
\begin{equation}
{\phi }_{\sq}(\xi )= \frac{([2]_{\sqrt{q}})^2 }{2} (-{\rm i})^{\sq}J_{\sq%
}(q, {\rm i}\sqrt{q}[2]_{\sqrt{q}}|\xi|) K_{\sq}(q,[2]_{\sqrt{q}}|\xi|).
\label{e27}
\end{equation}
Note that the integral over $\theta$ is a standard integration while that over $%
|\xi|$ is a $q$-integration.

In fact, for $\q\geq 0$, we have
\begin{eqnarray}
I_{\q} &=&\int \frac{d_q^2{\xi }}\pi {\phi }_{\q}(\xi )N_{\sq%
}^2\sum\limits_{j=0}^{k-1}\sum\limits_{n,m}^{}\frac{{\xi }^{kn+j}{{\xi }^{*}}%
^{km+j}|kn+j+\q,kn+j{\rangle }{\langle }km+j+\q,km+j|}{{\{[kn+j]![kn+j+%
\q]![km+j]![km+j+\q]!\}}^{1/2}}  \nonumber \\
&=&\int\limits_0^\infty \frac{d_{\sqrt{q}}|\xi |}\pi \frac{{([2]_{\sqrt{q}})}%
^2 }{2} {|\xi |}^{\sq+1} K_{\sq}(q,[2]_{\sqrt{q}}|\xi|)
\sum\limits_{j=0}^{k-1}\sum\limits_{n,m}^{}{|\xi |}^{k(n+m)+2j}\int\limits_{-\pi
}^\pi d{\theta }e^{{\rm i}k(n-m){\theta }}  \nonumber \\
&&\mbox{} {\times }\frac{|kn+j+\q,kn+j{\rangle }{\langle }km+j+\q,km+j|}{{%
\{[kn+j]![kn+j+\q]![km+j]![km+j+\q]!\}}^{1/2}}  \nonumber \\
&=& \int\limits_0^\infty {d_{\sqrt{q}}|\xi |} {([2]_{\sqrt{q}})}^2 {|\xi |}^{%
\q+1} K_{\sq}(q,[2]_{\sqrt{q}}|\xi|)
\nonumber \\
&& \mbox{} \times
\sum\limits_{j=0}^{k-1}\sum\limits_{n=0}^\infty \frac{{\ ({|\xi |}^2)}%
^{kn+j}|kn+j+\q,kn+j{\rangle }{\langle }kn+j+\q,kn+j|}{[kn+j]![kn+j+\q]!}
\nonumber \\
&=&\sum\limits_{n=0}^\infty \frac {{([2]_{\sqrt{q}})}^2 } {[n]![n+\q]!}|n+%
\q,n{\rangle }{\langle }n+\q,n|\int\limits_0^\infty {d_{\sqrt{q}}|\xi |%
}{|\xi |}^{2n+\q+1} K_{\q}(q,[2]_{\sqrt{q}}|\xi|)  \nonumber \\
&=&\sum\limits_{n=0}^\infty |n+\q,n{\rangle }{\langle }n+\q,n|.
\end{eqnarray}
Similarly, for $\q\leq 0$, we get
\begin{equation}
I_{\sq}=\sum\limits_{n=0}^\infty |n,n-\q{\rangle }{\langle }n,n-\q|.
\label{e29}
\end{equation}
Consequently, we derive
\begin{eqnarray}
\sum\limits_{\sq=-\infty }^\infty I_{\sq} &=&\sum\limits_{n=0}^\infty
\left( \sum\limits_{\sq=-\infty }^{-1}|n,n-\q{\rangle }{\langle
}n,n-\q|+\sum\limits_{\sq=0}^\infty |n+\q,n{\rangle }{\langle }n+\q%
,n|\right)  \nonumber \\
&=&\sum\limits_{m=0}^\infty \sum\limits_{n=0}^\infty
|m,n{\rangle }{\langle } m,n|=I.
\end{eqnarray}
Hence, the {\sl k}-component $q$-deformed charge coherent states are qualified to
make up an (over)complete representation. It should be mentioned that $I_{\sq}$
represents the resolution of unity in the subspace where $Q=\q$.

In the two special cases of $k=1$ and $k=2$, the above demonstration gives
the proof of completeness of the $q$-deformed charge coherent states [65]
and the even (odd) $q$-deformed charge coherent states [57], respectively.


\section*{\boldmath 4. Generation of {\sl k}-component $q$-deformed charge coherent states}

\mbox{}\hspace{6mm}The {\sl k}-component $q$-deformed coherent states,
defined as the $k$ orthonormalized eigenstates of the $k$th power of the
$q$-boson annihilation operator, can be expanded in the single-mode Fock
space as [58,59]
\begin{equation}
|{\xi },k{\rangle }_j=N_k^j\sum\limits_{n=0}^\infty \frac{{\xi }^{kn+j}}
{\sqrt{[kn+j]!} }|kn+j{\rangle },  \label{e31}
\end{equation}
where $j=0,1,\ldots ,k-1$ and
\begin{equation}
N_k^j{\equiv }N_k^j({|\xi |}^2)=
{ \left\{ \sum\limits_{n=0}^\infty \frac{{({|\xi |}^2)}^{kn+j}} {[kn+j]!}
\right\} }^{-1/2}.
\label{e32}
\end{equation}
As a special case, for $k=1$, $|{\xi },1{\rangle }_0$ are exactly the
$q$-deformed coherent states, i.e.,
\begin{equation}
|{\xi },1{\rangle }_0=e_q^{-{1/2}}({|\xi |}^2)
\sum\limits_{n=0}^\infty
\frac{{\xi }^n}{\sqrt{[n]!}}|n{\rangle }{\equiv }|{\xi }{\rangle }.
\label{e33}
\end{equation}

The {\sl k}-component $q$-deformed charge coherent states can also be
obtained from the states (31) and (33) according to the following expression
\begin{equation}
|\xi ,\q,k{\rangle }_{j}
=\left\{
\begin{array}{ll}
N_{k\sq}^{j}e_q^{{1/2}} ({|{\xi }_1|}^2) {\left[ N_{k}^{j}({|{\xi }_2|}%
^2)\right] }^{-1}{\xi _1}^{-\sq}\int\limits_{-\pi }^\pi \frac{d{\alpha }}{%
2\pi }e^{+{\rm i}\sq\alpha }|e^{-{\rm i}\alpha }{\xi }_1{\rangle }{\otimes
}|e^{{\rm i}\alpha }{%
\xi }_2,k{\rangle }_{j}, & \q\geq 0, \\
N_{k\sq}^{j} e_q^{{1/2}} ({|{\xi }_1|}^2) {\left[ N_{k}^{j}({|{\xi }_2|}%
^2)\right] }^{-1}{\xi _1}^{+\sq}\int\limits_{-\pi }^\pi \frac{d{\alpha }}{%
2\pi }e^{-{\rm i}\sq\alpha }|e^{{\rm i}\alpha }{\xi }_2,k{\rangle }_{j}{\otimes }%
|e^{-{\rm i}\alpha }{\xi }_1{\rangle }, & \q\leq 0,
\end{array}
\right.  \label{e43}
\end{equation}
where $\xi ={\xi }_1{\xi }_2$. Such a representation is very useful since
the properties of $q$-deformed coherent states and {\sl k}-component
$q$-deformed coherent states can now be employed in a study of the
properties of {\sl k}-component $q$-deformed charge coherent states. The
expression for the latter given in (34) has a very simple group-theoretical
interpretation: in (34) one suitably averages over the U(1)-group (caused by
the charge operator $Q$) action on the product of $q$-deformed coherent
states and {\sl k}-component $q$-deformed coherent states, which then
projects out the $Q=\q$ charge subspace contribution.

It is easy to see that in the limit $q{\rightarrow}1$, the above discussion
gives back the corresponding results for the usual {\sl k}-component charge
coherent states obtained in Ref. [37], and that in the two special cases of
$k=1$ and $k=2$, it gives the corresponding results for the $q$-deformed
charge coherent states [54] and the even (odd) $q$-deformed charge coherent
states [57], respectively.


\section*{\boldmath 5. $D$-algebra realization of SU$_q(1,1)$ generators}

\mbox{}\hspace{6mm}As is well known, the coherent state $D$-algebra [6,66]
is a mapping of quantum observables onto a differential form that acts on
the parameter space of coherent states, and has a beautiful application in
the reformulation of the entire laser theory in terms of $C$-number
differential equations [67]. We shall construct the $D$-algebra realization
of the $q$-deformed SU$_q(1,1)$ generators corresponding to the unnormalized
{\sl k}-component $q$-deformed charge coherent states, defined by
\begin{equation}
||\q{\rangle }_{j}{\equiv }||\xi ,\q,k{\rangle }_{j}=
{(N_{k\sq}^j)}^{-1}
|\xi ,\q,k{\rangle }_{j}.
\label{e35}
\end{equation}

Let $||\q{\rangle }$ denote a column vector composed of $||\q{\rangle }_j$
($j=0,1,\ldots ,k-1$), i.e.,
\begin{equation}
||\q{\rangle }{\equiv }\left[
\begin{array}{l}
||\q{\rangle }_0 \\
||\q{\rangle }_1 \\
\vdots \\
||\q{\rangle }_{k-1}
\end{array}
\right] .  \label{e36}
\end{equation}
The action of the operators $a_i$, $a_i^{+}$ and $N_i$ on this column vector
can be written in the matrix form:

\begin{equation}
\begin{array}{ll}
{\rm Positive\ } Q & {\rm Negative\ } Q \\[0.4cm]
a_1
||\q{\rangle }
=||\q-1{\rangle }, & a_1
||\q{\rangle }={\xi }M
||\q-1{\rangle }, \\[0.4cm]
a_2||\q{\rangle }
={\xi }M ||\q+1{\rangle }, & a_2
||\q{\rangle }=||\q+1{\rangle }, \\[0.4cm]
a_1^{+}
||\q{\rangle }
={\xi }^{-\sq}\frac d{d_q\xi }{\xi}^{\sq+1}
||\q+1{\rangle }, & a_1^{+}
||\q{\rangle }= \frac d{d_q\xi }N
||\q+1{\rangle }, \\[0.4cm]
a_2^{+}||\q{\rangle }
=\frac d{d_q\xi }N
||\q-1{\rangle }, & a_2^{+}
||\q{\rangle }
= {\xi }^{\sq}\frac d{d_q\xi }{\xi}^{-\sq+1}
||\q-1{\rangle }, \\[0.4cm]
N_1||\q{\rangle }
= \left({\xi }\frac d{d\xi }+\q \right)
||\q{\rangle }, & N_1
||\q{\rangle }
= \xi\frac d{d\xi }
||\q{\rangle }, \\[0.4cm]
N_2||\q{\rangle }
= {\xi }\frac d{d\xi }
||\q{\rangle }, & N_2
||\q{\rangle }
= \left(\xi\frac d{d\xi }-\b{q}\right)
||\q{\rangle },
\end{array}
\label{e37}
\end{equation}
where $d/d\xi$ is a standard differential operator, whereas
$d/d_q\xi$ is a $q$-differential one [42,47,64], defined by
\begin{equation}  \label{e38}
\frac{d}{d_q\xi}f(\xi)=\frac{f(q\xi)-f(q^{-1}\xi)}{q\xi-q^{-1}\xi };
\end{equation}
and
\begin{equation}  \label{e39}
M=\left[
\begin{array}{lllll}
0 & 0 & \cdots & 0 & 1 \\
1 & 0 & \cdots & 0 & 0 \\
0 & 1 & \cdots & 0 & 0 \\
\vdots & \vdots & \ddots & \vdots & \vdots \\
0 & 0 & \cdots & 1 & 0
\end{array}
\right],{\hskip 0.43in} N=\left[
\begin{array}{lllll}
0 & 1 & 0 & \cdots & 0 \\
0 & 0 & 1 & \cdots & 0 \\
\vdots & \vdots & \vdots & \ddots & \vdots \\
0 & 0 & 0 & \cdots & 1 \\
1 & 0 & 0 & \cdots & 0
\end{array}
\right].
\end{equation}
Obviously, $N$ is both the inverse and the transpose of M.

The $q$-deformed SU$_q(1,1)$ algebra consists of three generators $K_0$, $%
K_{+}$, and $K_{-}$, satisfying the commutation relations
\begin{equation}
\lbrack K_{+},K_{-}]=-[2K_0], \qquad [K_0,K_{\pm }]={\pm }K_{\pm },
\label{e40}
\end{equation}
and is realized in terms of the two-mode $q$-boson operators as
\begin{equation}
K_{-}=a_1a_2, \qquad K_{+}=a_1^{+}a_2^{+}, \qquad K_0={\frac
12}(N_1+N_2+1).  \label{e41}
\end{equation}
Actually, the {\sl k}-component $q$-deformed charge coherent states are also
the $k$ orthonormalized eigenstates of the $k$th power of $K_{-}$.

The $D$-algebra of the SU$_q(1,1)$ generators $A$ may be defined for the
action on the ket coherent states (36) or for that on the corresponding bras
as
\begin{eqnarray}
A||\q{\rangle }
&=&D^k(A)||\q{\rangle }, \label{e42}\\
{\langle }\q||A &=&D^b(A)
{\langle }\q||,
\end{eqnarray}
respectively.
Using (\ref{e37}) and (\ref{e41}), we get for the former
\begin{eqnarray}
D^k(K_{-})& = &{\xi }M, \label{e44} \\
D^k(K_{+}) & = & {\xi }^{-|\sq|}\frac d{d_q\xi }{\xi}^{|\sq|+1}
\frac d{d_q\xi }N, \\
D^k(K_0) & = &\frac 12 \left(2{\xi }\frac d{d\xi }+|\q|+1\right)I,
\end{eqnarray}
while the latter can be obtained from the adjoint relation
\begin{equation}
D^b(A)={\left[ D^k(A^{+})\right] }^{*}.  \label{e47}
\end{equation}
Thus, the $D$-algebra of the SU$_q(1,1)$ generators corresponding to the
unnormalized {\sl k}-component $q$-deformed charge coherent states has been
realized in a $q$-differential-operator matrix form.

From (36), (39), (42) and (44), we clearly see that by the successive
actions of the operator $a_1a_2$, each component of the {\sl k}-component
$q$-deformed charge coherent states, apart from normalization, can be
transformed into another in this way: $|\xi,\q,k{\rangle}_{0}\rightarrow
|\xi,\q,k{\rangle}_{k-1}\rightarrow |\xi,\q,k{\rangle}_{k-2}\rightarrow
\cdots\rightarrow|\xi,\q,k{\rangle}_{1}\rightarrow |\xi,\q,k{\rangle}_{0}$.
Actually, $a_1a_2$ plays the role of a rotating operator among
these {\sl k}-component states.

It is easy to check that in the limit $q{\rightarrow}1$, the above
discussion gives back the corresponding results for the usual
{\sl k}-component charge coherent states obtained in Ref. [37], and that in
the two special cases of $k=1$ and $k=2$, it gives the corresponding results
for the $q$-deformed charge coherent states [65] and the even (odd)
$q$-deformed charge coherent states [57], respectively.


\section*{\boldmath 6. Nonclassical properties of {\sl k}-component
$q$-deformed charge coherent states}

\mbox{}\hspace{6mm}In Ref. [57], the authors (X.-M.L. and C.Q.) have
examined the even and odd $q$-deformed charge coherent states for some
nonclassical properties, such as SU$_q(1,1)$ squeezing, single- or two-mode
$q$-squeezing, and two-mode $q$-antibunching. In this section, we will study
the nonclassical properties of the {\sl k}-component $q$-deformed charge
coherent states with $k\geq3$.


\subsection*{\boldmath 6.1. SU$_q(1,1)$ squeezing}

\mbox{}\hspace{6mm}In analogy with the definition of SU(1,1) squeezing [68],
we have introduced SU$_q(1,1)$ squeezing [57] in terms of the Hermitian
$q$-deformed quadrature operators
\begin{equation}
X_1=\frac{K_{+}+K_{-}}2, \qquad X_2=\frac{{\rm i}(K_{+}-K_{-})}2,  \label{e48}
\end{equation}
which satisfy the commutation relation
\begin{equation}
\lbrack X_1,X_2]=\frac{\rm i}{2}[2K_0]  \label{e49}
\end{equation}
and the uncertainty relation 
\begin{equation}
{\langle }{({\Delta }X_1)}^2{\rangle }{\langle }{({\Delta }X_2)}^2{\rangle }{%
\geq }\frac 1{16}{|{\langle }[2K_0]{\rangle }|}^2.  \label{e50}
\end{equation}
A state is said to be SU$_q(1,1)$ squeezed if
\begin{equation}
{\langle }{({\Delta }X_i)}^2{\rangle }<\frac 14{|{\langle }[2K_0]{\rangle }|}
\qquad (i=1 {\rm \ or\ }2).  \label{e51}
\end{equation}

Let us now calculate the fluctuations (variances) of $X_1$ and $X_2$ with
respect to the {\sl k}-component $q$-deformed charge coherent states. Using
(41) -- (44) and (47), we get
\begin{eqnarray}
&&_0{\langle }{\xi },\q,k|K_{+}K_{-}|\xi ,\q,k{\rangle }_0={|\xi |}^2{{%
(N_{k\sq}^0)}^2}/{{(N_{k\sq}^{k-1})}^2}, \\
&&_m{\langle }{\xi },\q,k|K_{+}K_{-}|\xi ,\q,k{\rangle }_m={|\xi |}^2{{%
(N_{k\sq}^m)}^2}/{{(N_{k\sq}^{m-1})}^2},{\hskip 0.22in}m=1,2,\ldots ,k-1.
\end{eqnarray}
In the meantime, for the states $|\xi ,\q,k{\rangle }_j$ $(k\geq 3)$, it
always follows that
\begin{equation}
_j{\langle }{\xi },\q,k|K_{-}|\xi ,\q,k{\rangle }_j={}_j{\langle }{\xi }%
,\q,k|K_{-}^2|\xi ,\q,k{\rangle }_j=0,{\hskip 0.30in}j=0,1,\ldots ,k-1.
\label{e54}
\end{equation}
Therefore, for $|\xi ,\q,k{\rangle }_0$ and $|\xi ,\q,k{\rangle }_m$~ $%
(m=1,2,\ldots ,k-1)$, the fluctuations are given by
\begin{eqnarray}
_0{\langle }{\xi },\q,k|{({\Delta }X_1)}^2|\xi ,\q,k{\rangle }_0 &=&{}_0{%
\langle }{\xi },\q,k|{({\Delta }X_2)}^2|\xi ,\q,k{\rangle }_0  \nonumber \\
&=&\frac 14{}_0{\langle }{\xi },\q,k|[2K_0]|\xi ,\q,k{\rangle }_0+\frac 12{|\xi |}%
^2{{(N_{k\sq}^0)}^2}/{{(N_{k\sq}^{k-1})}^2}, \\
_m{\langle }{\xi },\q,k|{({\Delta }X_1)}^2|\xi ,\q,k{\rangle }_m &=&{}_m{%
\langle }{\xi },\q,k|{({\Delta }X_2)}^2|\xi ,\q,k{\rangle }_m  \nonumber \\
&=&\frac 14{}_m{\langle }{\xi },\q,k|[2K_0]|\xi ,\q,k{\rangle }_m+\frac 12{|\xi |}%
^2{{(N_{k\sq}^m)}^2}/{{(N_{k\sq}^{m-1})}^2}.~~~~
\end{eqnarray}
Consequently, for $k\geq 3$, we find
\begin{eqnarray}
_j{\langle }{\xi },\q,k|{({\Delta }X_1)}^2|\xi ,\q,k{\rangle }_j &=&{}_j{%
\langle }{\xi },\q,k|{({\Delta }X_2)}^2|\xi ,\q,k{\rangle }_j  \nonumber \\
&\geq &\frac 14{}_j{\langle }{\xi },\q,k|[2K_0]|\xi ,\q,k{\rangle }_j,{\hskip %
0.26in}j=0,1,\ldots ,k-1.
\end{eqnarray}

The inequalities in (57) say that there is no SU$_q(1,1)$ squeezing in the
{\sl k}-component $q$-deformed charge coherent states with $k\geq3$. However,
there is such squeezing in the even and odd $q$-deformed charge coherent
states [57].

It is easy to verify that the $q$-deformed charge coherent states satisfy
the equality in (50) and that
${\langle}{({\Delta}X_{1})}^2{\rangle}={%
\langle}{({\Delta}X_{2})}^2{\rangle}$.
This point has been observed in Ref. [57].
Therefore, the $q$-deformed charge coherent states are not SU$_q(1,1)$
squeezed.


\subsection*{\boldmath 6.2. Single-mode $q$-squeezing}

\mbox{}\hspace{6mm}In analogy with the definition of single-mode squeezing
[27], we have introduced single-mode $q$-squeezing [57] in terms of the
Hermitian $q$-deformed quadrature operators for the individual modes
\begin{eqnarray}
Y_1 &=&\frac{a_1^{+}+a_1}2,\qquad Y_2=\frac{{\rm i}(a_1^{+}-a_1)}2,
\nonumber \\
Z_1 &=&\frac{a_2^{+}+a_2}2,\qquad Z_2=\frac{{\rm i}(a_2^{+}-a_2)}2,
\end{eqnarray}
which satisfy the commutation relations
\begin{equation}
\lbrack Y_1,Y_2]=\frac {\rm i}2[a_1,a_1^+],\qquad [Z_1,Z_2]=\frac
{\rm i}2[a_2,a_2^+],  \label{e59}
\end{equation}
and the uncertainty relations
\begin{equation}
{\langle }{({\Delta }Y_1)}^2{\rangle }{\langle }{({\Delta }Y_2)}^2{\rangle }{%
\geq }\frac 1{16} {|{\langle }[a_1,a_1^+]{\rangle }|}^2,\qquad {%
\langle }{({\Delta }Z_1)}^2{\rangle }{\langle }{({\Delta }Z_2)}^2{\rangle }{%
\geq }\frac 1{16}{|{\langle }[a_2,a_2^+]{\rangle }|}^2.  \label{e60}
\end{equation}
A state is said to be single-mode $q$-squeezed if
\begin{equation}
{\langle }{({\Delta }Y_i)}^2{\rangle }<\frac 14 |{\langle }[a_1,a_1^+]{%
\rangle }|,\qquad {\langle }{({\Delta }Z_i)}^2{\rangle }<\frac 14 |{%
\langle }[a_2,a_2^+]{\rangle }|~\hspace{5mm}(i=1{\rm \ or\ }2).  \label{e61}
\end{equation}

For the states $|\xi,\q,k{\rangle}_j$ $(k\geq1)$, it always follows that
\begin{equation}  \label{e62}
_{j}{\langle}a_1{\rangle}_{j}={}_j{\langle}a_2{\rangle}_{j}={}_{j}{\langle}a_1^2{%
\rangle}_{j}={}_{j}{\langle}a_2^2{\rangle}_{j}={}_{j}{\langle}a_1^+a_2{\rangle}%
_{j}=0,{\hskip 0.30in} j=0,1,\ldots,k-1.
\end{equation}
Thus, the fluctuations are given by
\begin{eqnarray}
_j{\langle}{\xi },\q,k|{({\Delta }Y_1)}^2|\xi ,\q,k{\rangle}_j&=& {}_j{\langle}%
{\xi },\q,k|{({\Delta }Y_2)}^2|\xi ,\q,k{\rangle}_j  \nonumber \\
&=&\frac {1}{4}\left\{  {}_j{\langle}{\xi },\q,k|[a_1,a_1^+]|\xi ,\q,k{\rangle}_j +2\,{}_j{\langle}{\xi },\q,k|a_1^+a_1|\xi ,\q,k{\rangle}%
_j\right\}  \nonumber \\
&>&\frac{1}{4}{}\,_j{\langle}{\xi },\q,k|[a_1,a_1^+]|\xi ,\q,k{\rangle}_j,
\label{63}\\
_j{\langle}{\xi },\q,k|{({\Delta }Z_1)}^2|\xi ,\q,k{\rangle}_j&=& {}_j{\langle}%
{\xi },\q,k|{({\Delta }Z_2)}^2|\xi ,\q,k{\rangle}_j  \nonumber \\
&=&\frac {1}{4}\left\{  {}_j{\langle}{\xi },\q,k|[a_2,a_2^+]|\xi ,\q,k{\rangle}_j +2\,{}_j{\langle}{\xi },\q,k|a_2^+a_2|\xi ,\q,k{\rangle}%
_j\right\} \nonumber \\
&>&\frac{1}{4}\,{}_j{\langle}{\xi },\q,k|[a_2,a_2^+]|\xi ,\q,k{\rangle}_j.
\label{64}
\end{eqnarray}
This shows that there is no single-mode $q$-squeezing in the
{\sl k}-component $q$-deformed charge coherent states with $k\geq1$. As two
special cases, there is no such $q$-squeezing in the $q$-deformed charge
coherent states [65] and the even (odd) $q$-deformed charge coherent states
[57] as $k$ becomes 1 and 2, respectively.


\subsection*{\boldmath 6.3. Two-mode $q$-squeezing}

\mbox{}\hspace{6mm}In analogy with the definition of two-mode squeezing
[69], we have introduced two-mode $q$-squeezing [57] in terms of the
Hermitian $q$-deformed quadrature operators for the two modes
\begin{equation}  \label{e65}
W_1=\frac{Y_1+Z_1}{\sqrt{2}}=\frac{1}{\sqrt{8}}(a_1^{+}+a_2^{+}+a_1+a_2),\quad
W_2=\frac{Y_2+Z_2}{\sqrt{2}}=\frac{\rm i}{\sqrt{8}}(a_1^{+}+a_2^{+}-a_1-a_2),
\end{equation}
which satisfy the commutation relation
\begin{equation}  \label{e66}
[W_1,W_2]=\frac{1}{4}{\rm i}\left\{[a_1,a_1^+]+[a_2,a_2^+]\right\}
\end{equation}
and the uncertainty relation
\begin{equation}  \label{e67}
{\langle}{({\Delta}W_{1})}^2{\rangle}{\langle}{({\Delta}W_{2})}^2{\rangle}{%
\geq}\frac{1}{64} {|{\langle }[a_1,a_1^+]{\rangle }+{\langle }[a_2,a_2^+]{%
\rangle } |}^2.
\end{equation}
A state is said to be two-mode $q$-squeezed if
\begin{equation}  \label{e68}
{\langle}{({\Delta}W_{i})}^2{\rangle}<\frac{1}{8} {|{\langle }[a_1,a_1^+]{%
\rangle }+{\langle }[a_2,a_2^+]{\rangle } |} ~ \hspace{8mm}(i=1 {\rm \ or\ }2).
\end{equation}

For the states $|\xi,\q,k{\rangle}_j$ $(k\geq2)$, the fluctuations are given by
\begin{eqnarray}
_j{\langle}{\xi},\q,k|{({\Delta }W_1)}^2|{\xi},\q,k{\rangle}_j&=& {}_j{\langle}%
{\xi},\q,k|{({\Delta }W_2)}^2|{\xi},\q,k{\rangle}_j  \nonumber \\
&=&\frac 12 \left\{ {}_j{\langle}{\xi},\q,k|{({\Delta }Y_1)}^2|{\xi},\q,k{\rangle}%
_j+ {}_j{\langle}{\xi},\q,k|{({\Delta }Z_1)}^2|{\xi},\q,k{\rangle}_j\right\}
\nonumber \\
&=&\frac 12 \left\{ {}_j{\langle}{\xi},\q,k|{({\Delta }Y_2)}^2|{\xi},\q,k{\rangle}%
_j+ {}_j{\langle}{\xi},\q,k|{({\Delta }Z_2)}^2|{\xi},\q,k{\rangle}_j\right\}
\nonumber \\
&=&\frac {1}{8}\left\{  {}_j{\langle}{\xi },\q,k|[a_1,a_1^+]|\xi ,\q,k{\rangle}_j +
{}_j{\langle}{\xi },\q,k|[a_2,a_2^+]|\xi ,\q,k{\rangle}_j
\right.
\nonumber \\
&&\mbox{} \left. +2\,{}_j{\langle}{\xi },\q,k|a_1^+a_1|\xi ,\q,k{\rangle}_j
+2\,{}_j{\langle}{\xi },\q,k|a_2^+a_2|\xi ,\q,k{\rangle}_j
\right\}  \nonumber \\
&>&\frac{1}{8}\left\{ {}_j{\langle}{\xi },\q,k|[a_1,a_1^+]|\xi ,\q,k{\rangle}_j
+ {}_j{\langle}{\xi },\q,k|[a_2,a_2^+]|\xi ,\q,k{\rangle}_j
\right\}.~~~~~
\label{69}
\end{eqnarray}
This shows that there is no two-mode $q$-squeezing in the
{\sl k}-component $q$-deformed charge coherent states with $k\geq2$.
As a special case, there is no such $q$-squeezing in the even and odd
$q$-deformed charge coherent states [57] as $k$ becomes 2.
However, there is such $q$-squeezing in the $q$-deformed charge coherent
states [65].


\subsection*{\boldmath 6.4. Two-mode $q$-antibunching}

\mbox{}\hspace{6mm}In analogy with the definition of two-mode antibunching
[36], we have introduced a two-mode $q$-correlation function as [57]
\begin{equation}
g^{(2)}(0)\equiv \frac{{\langle }{(a_1^{+}a_2^{+})}^2{(a_1a_2)}^2{\rangle }}{{{%
\langle }a_1^{+}a_2^{+}a_1a_2{\rangle }}^2}=\frac{{\langle }:{([N_1][N_2])}%
^2:{\rangle }}{{{\langle }[N_1][N_2]{\rangle }}^2}
=\frac{ {\langle }K_{+}^2K_{-}^2{\rangle }  } { { {\langle }
K_{+}K_{-} {\rangle } }^2 }, \label{e70}
\end{equation}
where $a_i$ and $a_i^{+}$ represent the annihilation and creation operators
of $q$-deformed photons of a deformed light field and $:\,:$ denotes normal
ordering. We call $g^{(2)}(0)$ the two-mode $q$-correlation degree.
Physically, $g^{(2)}(0)$ is a measure of $q$-deformed two-photon
correlations in the $q$-deformed two-mode field and is related to the $q$%
-deformed two-photon number distributions. A state is said to be two-mode $%
q$-antibunched if
\begin{equation}
g^{(2)}(0)<1.  \label{e71}
\end{equation}

Let us now study the two-mode $q$-antibunching effect for the
{\sl k}-component $q$-deformed charge coherent states with $k\geq3$ . First,
for $k\geq3$, one can easily obtain the following relations:
\begin{eqnarray}
&&_0{\langle}{\xi },\q,k|K_{+}^2K_{-}^2|\xi ,\q,k{\rangle}_0={|\xi |}^4 {{%
(N_{k\sq}^0)}^2}/{{(N_{k\sq}^{k-2})}^2}, \\
&&_1{\langle}{\xi },\q,k|K_{+}^2K_{-}^2|\xi ,\q,k{\rangle}_1={|\xi |}^4 {{%
(N_{k\sq}^1)}^2}/{{(N_{k\sq}^{k-1})}^2}, \\
&&_l{\langle}{\xi },\q,k|K_{+}^2K_{-}^2|\xi ,\q,k{\rangle}_l={|\xi |}^4 {{%
(N_{k\sq}^l)}^2}/{{(N_{k\sq}^{l-2})}^2},{\hskip 0.22in} l=2,3,\ldots,k-1.
\end{eqnarray}
According to (52), (53) and (72) -- (74),
the two-mode $q$-correlation degrees
of the {\sl k}-component $q$-deformed charge coherent states can be obtained
as follows:
\begin{eqnarray}
&&g_0^{(2)}(0)=\frac {_0{\langle}{\xi },\q,k|K_{+}^2K_{-}^2|\xi ,\q,k{\rangle}%
_0} {{( {_0{\langle}{\xi },\q,k|K_{+}K_{-}|\xi ,\q,k{\rangle}_0})}^2 } =\frac {%
{(N_{k\sq}^{k-1})}^4 } {{(N_{k\sq}^{0})}^2 {(N_{k\sq}^{k-2})}^2 }, \\
&&g_1^{(2)}(0)=\frac {_1{\langle}{\xi },\q,k|K_{+}^2K_{-}^2|\xi ,\q,k{\rangle}%
_1} {{( {_1{\langle}{\xi },\q,k|K_{+}K_{-}|\xi ,\q,k{\rangle}_1})}^2} =\frac {{%
(N_{k\sq}^{0})}^4 } {{(N_{k\sq}^{1})}^2 {(N_{k\sq}^{k-1})}^2 }, \\
&&g_l^{(2)}(0)=\frac {_l{\langle}{\xi },\q,k|K_{+}^2K_{-}^2|\xi ,\q,k{\rangle}%
_l} {{({_l{\langle}{\xi },\q,k|K_{+}K_{-}|\xi ,\q,k{\rangle}_l})}^2} =\frac {{%
(N_{k\sq}^{l-1})}^4 } {{(N_{k\sq}^{l-2})}^2 {(N_{k\sq}^{l})}^2 },{\hskip 0.10in}
l=2,3,\ldots,k-1.
\end{eqnarray}
Evidently, the following relation exists:
\begin{equation}  \label{e78}
\prod\limits_{j=0}^{k-1}g_j^{(2)}(0)=1.
\end{equation}
We shall prove that for $k\geq3$, the {\sl k}-component $q$-deformed charge
coherent states show two-mode $q$-antibunching.

From (12) and (75), it follows that
\begin{eqnarray*}
g_0^{(2)}(0)=
\frac{f(x)}{x^k{\varphi}(x)},
\end{eqnarray*}
where
\begin{eqnarray*}
f(x)={ \sum\limits_{m=0}^{\infty}\left\{\sum\limits_{n=0}^{m}
\frac{1}{[kn]![kn+|\q|]![km-kn+k-2]![km-kn+k-2+|\q|]!}\right\} x^{km}},
\end{eqnarray*}
\begin{eqnarray*}
{\varphi}(x)={\sum\limits_{m=0}^{\infty} \left\{\sum\limits_{n=0}^{m}\frac{1} {%
[kn+k-1]![kn+k-1+|\q|]![km-kn+k-1]![km-kn+k-1+|\q|]!}\right\} x^{km}}
\end{eqnarray*}
and $x={|\xi|}^2$. For $k\geq3$, we have
\begin{eqnarray*}
&&\sum\limits_{n=0}^{m} \frac{1}{[kn]![kn+|\q|]![km-kn+k-2]![km-kn+k-2+|\q|]!}
\\
&& \ \ \ \ {>} \sum\limits_{n=0}^{m} \frac{1} {%
[kn+k-1]![kn+k-1+|\q|]![km-kn+k-1]![km-kn+k-1 +|\q|]!},
\end{eqnarray*}
and thus $f(x)>\varphi(x)$ when $x>0$. Hence, $g_0^{(2)}(0)>1$ when
$0<x\leq1$. However, when $x>1$, the following inequality
\begin{eqnarray*}
\frac{f(x)}{x^k{\varphi}(x)}<1,{\rm \ i.e.,\ }x^{k}> {\frac{{f(x)}}{{{\varphi}(x)}}}
\end{eqnarray*}
may have real roots. Consequently, in the region of $x>1$,
for arbitrary fixed values of $\q$ and $q$, there surely exists some range
of $x$ values such that
\begin{eqnarray*}
g_0^{(2)}(0)=\frac{f(x)}{x^k{\varphi}(x)}<1.
\end{eqnarray*}
To make the above statement clear, we plot $g_0^{(2)}(0)$ against $x$ for
various $k$, $\q$ and $q$ in Fig. 1.

From (12) and (76), it follows that
\begin{eqnarray*}
g_1^{(2)}(0)=
\frac{x^kf_1(x)}{{%
\varphi}_1(x)},
\end{eqnarray*}
where
\begin{eqnarray*}
f_1(x)={\sum\limits_{m=0}^{\infty}\left\{\sum\limits_{n=0}^{m}
\frac{1}{[kn+1]![kn+1+|\q|]![km-kn+k-1]![km-kn+k-1+|\q|]!}\right\} x^{km}},
\end{eqnarray*}
\begin{eqnarray*}
{\varphi}_1(x)=
{\sum\limits_{m=0}^{\infty} \left\{\sum\limits_{n=0}^{m}\frac{1} {%
[kn]![kn+|\q|]![km-kn]![km-kn+|\q|]!}\right\} x^{km}}.
\end{eqnarray*}
Apparently,
\begin{eqnarray*}
&&\sum\limits_{n=0}^{m} \frac{1}{[kn+1]![kn+1+|\q|]![km-kn+k-1]![km-kn+k-1+|\q|]!%
} \\
&& \ \ \ \ {<}\sum\limits_{n=0}^{m}\frac{1} {[kn]![kn+|\q|]![km-kn]![km-kn+|\q|]!},
\end{eqnarray*}
so that $f_1(x)<{\varphi}_1(x)$. Therefore, $g_1^{(2)}(0)<x^k$, namely,
$g_1^{(2)}(0)<1$ as $x\leq1$.

From (12) and (77), it follows that
\begin{eqnarray*}
g_l^{(2)}(0)=
\frac{f_2(x)}{{%
\varphi}_2(x)},
\end{eqnarray*}
where
\begin{eqnarray*}
f_2(x)=
{\sum\limits_{m=0}^{\infty}\left\{\sum\limits_{n=0}^{m}
\frac{1}{[kn+l-2]![kn+l-2+|\q|]![km-kn+l]![km-kn+l+|\q|]!}\right\} x^{km}},
\end{eqnarray*}
\begin{eqnarray*}
{\varphi}_2(x)=
{\sum\limits_{m=0}^{\infty} \left\{\sum\limits_{n=0}^{m}\frac{1} {%
[kn+l-1]![kn+l-1+|\q|]![km-kn+l-1]![km-kn+l-1+|\q|]!}\right\} x^{km}}.
\end{eqnarray*}
Obviously,
\begin{eqnarray*}
f_2(x)&{<}&\frac{1}{[l-2]![l-2+|\q|]![l]![l+|\q|]!}
\sum\limits_{m=0}^{\infty}(m+1)x^{km}, \\
{\varphi}_2(x)&{>}&
\sum\limits_{m=0}^{\infty}    \frac{(m+1)x^{km} }
{ {\{[km+l-1]![km+l-1+|\q|]!\}}^2 }
~~{>}~~
\frac{1}{{\{[l-1]![l-1+|\q|]!\}}^2}.
\end{eqnarray*}
Thus, we obtain
\begin{eqnarray*}
g_l^{(2)}(0)
{<}\frac{{\{[l-1]![l-1+|\q|]!\}}^2}
{[l-2]![l-2+|\q|]![l]![l+|\q|]!}
\sum\limits_{m=0}^{\infty}(m+1)x^{km}.
\end{eqnarray*}
For $x<1$, it reads
\begin{eqnarray*}
\sum\limits_{m=0}^{\infty}(m+1)x^{km}=\frac{1}{{(1-x^k)}^2}.
\end{eqnarray*}
Therefore, as $x<1$, we get
\begin{eqnarray*}
g_l^{(2)}(0)<\frac{[l-1][l-1+|\q|]}{[l][l+|\q|]}\frac{1}{{(1-x^k)}^2}.
\end{eqnarray*}
As a result,
if $x^k\leq1-{\{[l-1][l-1+|\q|]/[l][l+|\q|]\}}^{1/2}$, then
\begin{eqnarray*}
g_l^{(2)}(0)<1,{\hskip 0.22in}l=2,3,\cdots,k-1.
\end{eqnarray*}

From the above discussion, we see that for $k\geq3$, the two-mode
$q$-correlation degrees $g_j^{(2)}(0)$ $(j =0, 1, \cdots, k-1)$ can be less
than 1 over some particular range of $x$ values.
This indicates that two-mode $q$-antibunching exists for the
{\sl k}-component $q$-deformed charge coherent states with $k\geq3$. The
same situation occurs for the even and odd $q$-deformed charge coherent
states [57]. However, for the  $q$-deformed charge coherent states we have
$g^{(2)}(0)=1$ so that no two-mode $q$-antibunching exists.

It can be shown that in the limit $q{\rightarrow }1$, the nonclassical
properties of the usual {\sl k}-component charge coherent states, studied
in Ref. [37], are retrieved as expected.


\section*{7. Summary}

\mbox{}\hspace{6mm}Let us sum up the results obtained in the present paper:

(1) The {\sl k}-component $q$-deformed charge coherent states, defined as
the $k$ $(k\geq 1)$ orthonormalized eigenstates of both the $k$th power of
the pair $q$-boson annihilation operator and the charge operator, have been
constructed and their (over)completeness proved. Such $q$-deformed states
become the usual {\sl k}-component charge coherent states in the limit
$q{\rightarrow}1$. They become the $q$-deformed charge coherent states and
the even (odd) $q$-deformed charge coherent states in the two special cases
of $k=1$ and $k=2$, respectively.

(2) The {\sl k}-component $q$-deformed charge coherent states have been shown
to be generated by a suitable average over the $U(1)$-group (caused by the
charge operator) action on the product of $q$-deformed coherent states and
{\sl k}-component $q$-deformed coherent states.

(3) The $D$-algebra of the SU$_q(1,1)$ generators corresponding to the
{\sl k}-component $q$-deformed charge coherent states has been realized in a
$q$-differential-operator matrix form.

(4) For $k\geq3$, the {\sl k}-component $q$-deformed charge coherent states
have been shown to exhibit two-mode $q$-antibunching, but neither
SU$_q$(1,1) squeezing, nor single- or two-mode $q$-squeezing.


\section*{Acknowledgments}

\mbox{}\hspace{6mm}X.-M.L.\ is grateful to Professor C. Quesne for warm
hospitality at the Universit\'{e} Libre de Bruxelles. He also acknowledges
financial support by the National Fund for Scientific Research (FNRS), Belgium, as well as
support by the National Natural Science Foundation of China under Grants Nos.
10174007, 10074008 and 60278021. C.Q.\ is a Research Director of the National Fund
for Scientific Research (FNRS), Belgium.


\newpage

\section*{Figure caption}

\mbox{}\hspace{6mm}Fig. 1. $g~({\equiv}g_0^{(2)}(0))$ against $x$ for
$k=3,~4,~5$,
with (a) $\q={\pm}2$, $q=0.9$ and (b) $\q={\pm}3$, $q=0.8$.

\newpage
\begin{figure}
\centering
\includegraphics[width=1.00\textwidth]{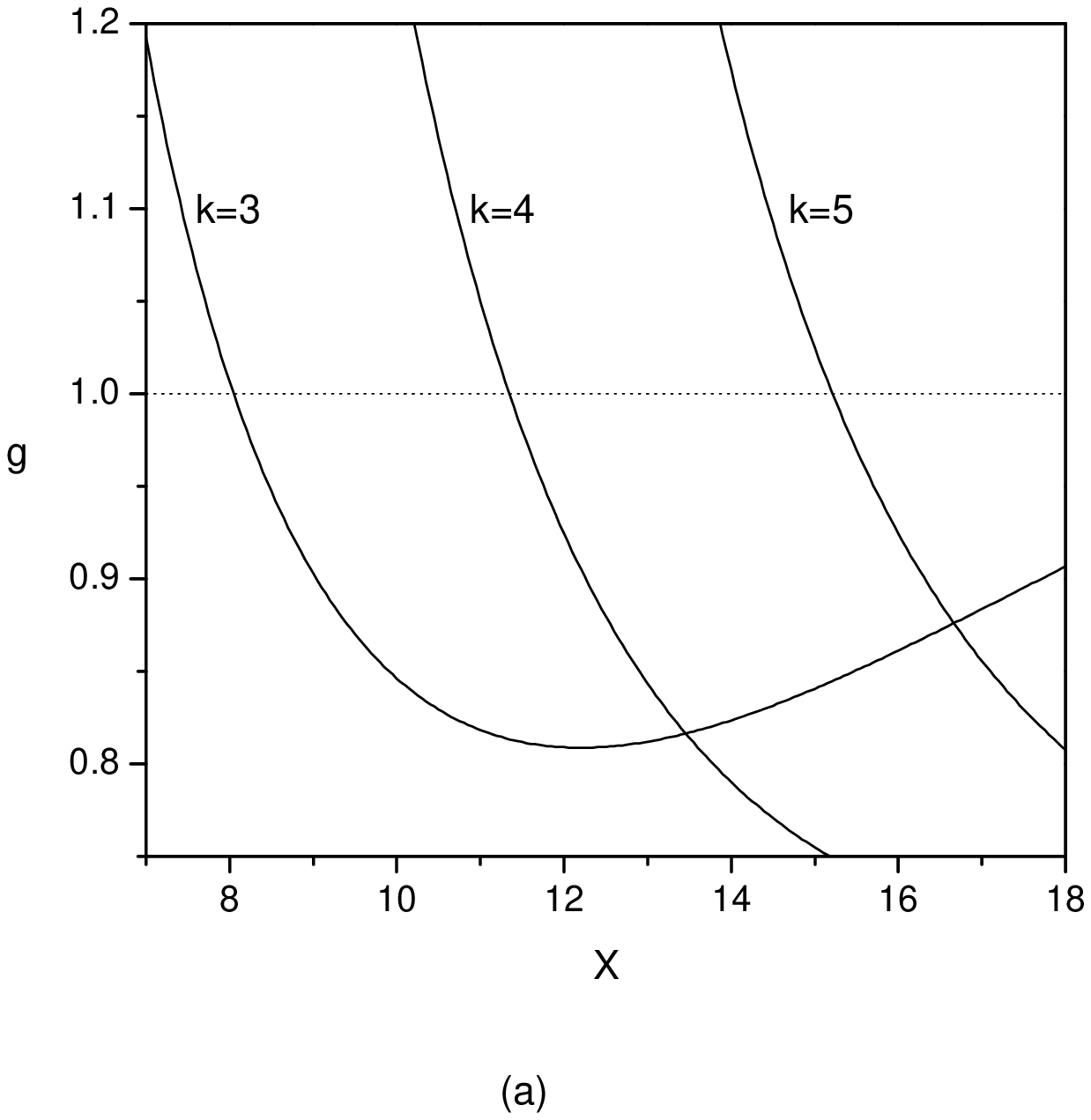}
\end{figure}

\newpage
\begin{figure}
\centering
\includegraphics[width=1.00\textwidth]{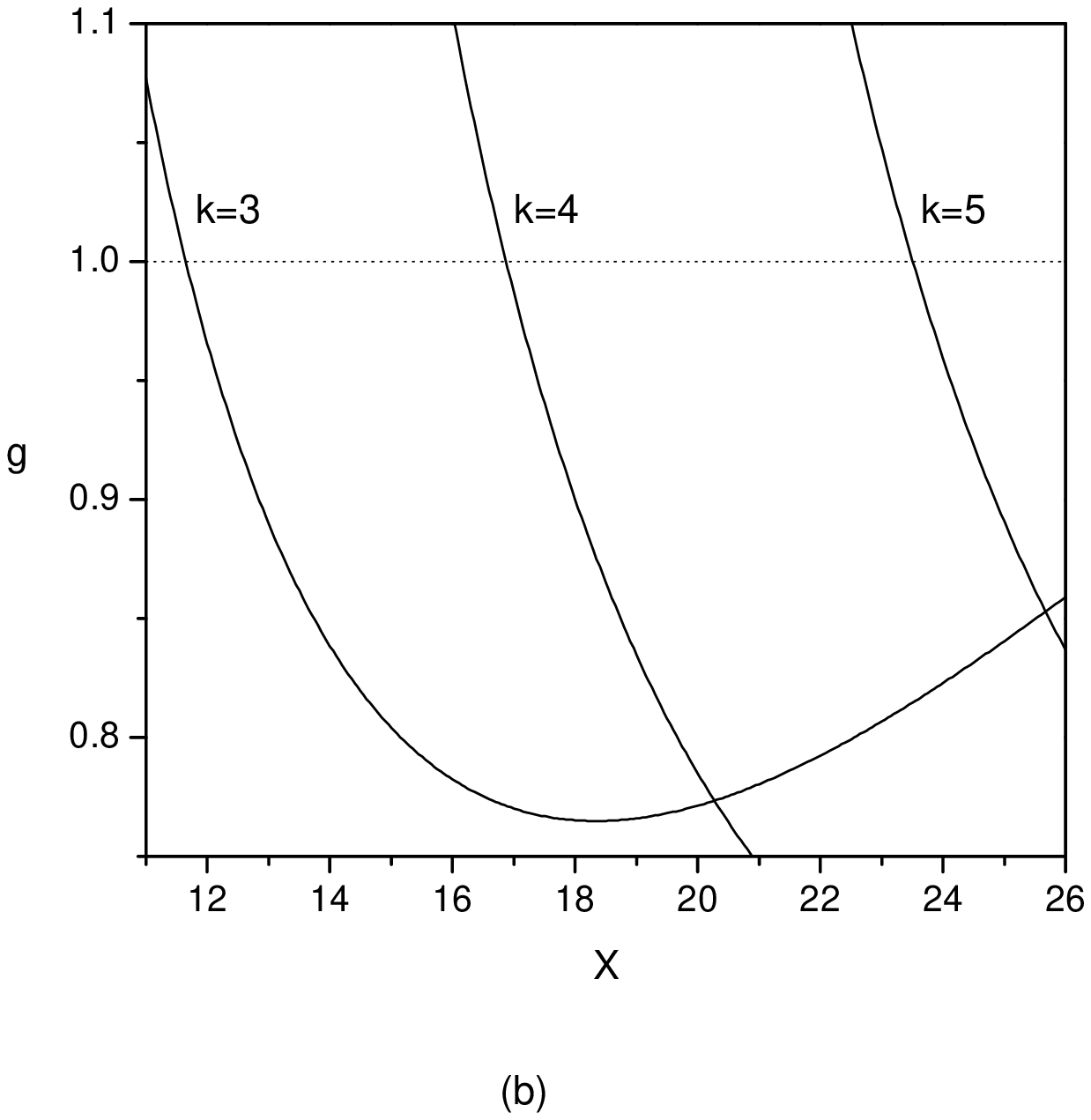}
\end{figure}

\end{document}